\documentclass[%
 reprint,
nofootinbib,
 amsmath,amssymb,
 aps,
 prd,
]{revtex4-2}

\usepackage{graphicx}
\usepackage{dcolumn}
\usepackage{bm}
\usepackage{hyperref}
\usepackage{physics}
\usepackage{subcaption}

\begin{document}

\title{Testable Flavored TeV-scale Resonant Leptogenesis with MeV-GeV Dark Matter in a Neutrinophilic 2HDM}

\author{Peisi Huang$^1$}
\email{peisi.huang@unl.edu}
\author{Kairui Zhang$^2$}
\email{kzhang25@ou.edu}
\affiliation{%
    $^1$Department of Physics and Astronomy, University of Nebraska, Lincoln, NE 68588, USA\\
    $^2$Homer L. Dodge Department of Physics and Astronomy, University of Oklahoma, Norman, OK 73019, USA
}%

\begin{abstract}
    We explore flavored resonant leptogenesis embedded in a neutrinophilic 2HDM. Successful leptogenesis is achieved by the very mildly degenerate two heavier right-handed neutrinos~(RHNs) $N_2$ and $N_3$ with a level of only $\Delta M_{32}/M_2 \sim \mathcal{O}(0.1\%-1\%)$. The lightest RHN, with a MeV–GeV mass, lies below the sphaleron freeze-out temperature and is stable, serving as a dark matter candidate. The model enables TeV-scale leptogenesis while avoiding the extreme mass degeneracy typically plagued conventional resonant leptogenesis. Baryon asymmetry, neutrino masses, and potentially even dark matter relic density can be addressed within a unified, experimentally testable framework.
\end{abstract}

\maketitle


\section{Introduction}

The observed baryon asymmetry of the universe~(BAU), $\eta_B \equiv \frac{n_B - n_{\bar{B}}}{n_\gamma}\bigg|_0 \approx 6\times 10^{-10}$ \cite{Cooke:2017cwo, Planck:2018vyg, Yeh:2022heq}, poses a significant challenge to the Standard Model~(SM). While the SM satisfies the Sakharov conditions for baryogenesis \cite{Sakharov:1967dj, Kuzmin:1985mm}, it fails to generate sufficient asymmetry. In particular, more baryon number violation, more CP violation, and a departure from thermal equilibrium unlike the one in the SM are needed. Traditional thermal leptogenesis within the type-I seesaw model introduces heavy RHNs whose CP-violating, out-of-equilibrium decays generate a lepton asymmetry, which sphalerons partially convert to baryon asymmetry~\cite{Fukugita:1986hr, Davidson:2008bu}. However, the Davidson-Ibarra bound, valid for hierarchical RHNs, requires the lightest RHN to have a mass of $M_1 \gtrsim 10^9$~GeV for sufficient asymmetry to be generated~\cite{Davidson:2002qv, Buchmuller:2004nz, Davidson:2008bu}. This heavy scale exacerbates the hierarchy problem of the SM~\cite{Vissani:1997ys} and introduces cosmological challenges, such as the gravitino problem in supersymmetric extensions~\cite{Kawasaki:2004qu, Ellis:1984eq}. Furthermore, this scale is far beyond the reach of any foreseeable experiments, making experimental verification difficult, if not impossible.

Resonant leptogenesis alleviates these issues by enhancing CP asymmetry through the interference between nearly degenerate RHN states~\cite{Pilaftsis:2003gt, Hambye:2004jf, DeSimone:2007edo, Pilaftsis:2004xx, Pilaftsis:2005rv,BhupalDev:2014pfm,BhupalDev:2014oar}. In this scenario, the self-energy contribution to CP asymmetry becomes resonantly enhanced when the mass splitting between two RHNs is comparable to their decay widths, $\Delta M \sim \Gamma$. This allows successful leptogenesis at TeV scales. However, achieving such near-degeneracy requires extreme fine-tuning of the RHN masses, $\Delta M/M \sim 10^{-9}$, which is viewed as unnatural.

The neutrinophilic Two-Higgs Doublet Model~($\nu$2HDM) offers an alternative by extending the SM+RHNs with a second Higgs doublet having a small vacuum expectation value~(VEV) and coupling exclusively to neutrinos. The smaller VEV reduces the equilibrium neutrino masses, increasing the CP asymmetry and allowing leptogenesis to occur at lower scales~\cite{Ma:2000cc, Haba:2011yc, Haba:2013pca, Clarke:2015hta}. However, $\nu$2HDM often suffers from strong $\Delta L = 2$ washout processes. Together with the Davidson-Ibarra bound, which still applies to these hierarchical leptogenesis models with a modified VEV, they impose a lower limit on the RHN mass of $M_N \gtrsim 10^5$GeV~\cite{Haba:2011yc}, making experimental verification still challenging.

Another drawback of most thermal leptogenesis models is that, contrary to one of the objectives of the type-I seesaw mechanism, they do not provide a dark matter~(DM) candidate, as they rely on the decay of the lightest RHN, $N_1$, to produce the asymmetry.

To address these challenges, this paper proposes a model combining the approaches of resonant leptogenesis and $\nu$2HDM. Inspired by neutrino mass pattern model building, we also introduce a RHN mass hierarchy: $M_1 \ll T_\text{spha} < M_2 \sim M_3$. Upon accounting for flavor effects, $N_2$ and $N_3$ enable TeV-scale resonant leptogenesis with only mild mass degeneracy breaking, $\Delta M_{32}/M_2 \sim \mathcal{O}(0.1\%-1\%)$. If this mass pattern arises from a fundamental symmetry breaking, the breaking of degeneracy directly relates to the scale of $N_1$~\cite{Merle:2012ya}, implying $M_1 \sim \mathcal{O}(\text{MeV-GeV}) \ll T_\text{spha}$ in our scenario. Since sphalerons remain in equilibrium until $T = 131.7$~GeV~\cite{DOnofrio:2014rug}, the lightest $N_1$ decouples from thermal leptogenesis and is stable. Thus, the model potentially addresses both BAU and the DM relic density, a feature absent in most thermal leptogenesis models. Furthermore, all model parameters lie in experimentally testable range, ensuring verification in the near future.

\section{Neutrino Masses}

We extend the SM by a second Higgs doublet. The Higgs sector couples to SM fermions in a Type-I 2HDM-like manner. We also introduce RHNs, which couple exclusively to the second Higgs doublet $\Phi_2$. Both $\Phi_1$ and $\Phi_2$ have hypercharge $+1$. A discrete $Z_2$ symmetry is imposed to suppress flavor-changing neutral currents at tree level, with $\Phi_1$ and SM fermions being $Z_2$-even, while $\Phi_2$ and the RHNs are $Z_2$-odd. Such scheme is often dubbed as $\nu$2HDM~\cite{Ma:2000cc, Haba:2011yc, Haba:2013pca, Clarke:2015hta}.

In this framework, $\Phi_1$ generates masses for SM fermions, while $\Phi_2$, with a small VEV $v_2$, facilitates the seesaw mechanism for neutrino masses. The SM VEV is $v = \sqrt{v_1^2 + v_2^2} \approx 246$~GeV, and we consider $v_2 \ll v_1$ such that $\tan\beta = v_1/v_2 \gg 1$. If one starts from other types of 2HDM, such as Type II, X, and Y, the Yukawa couplings hit a Landau pole for large $\tan\beta$, limiting the parameter space~\cite{Bijnens:2011gd}. The $\nu$2HDM avoids this issue and remains perturbative up to the Planck scale. The Yukawa interactions are described by
\begin{align}
    -\mathcal{L}_Y = \ &y_u \overline{Q}_L \tilde{\Phi}_1 u_R + y_d \overline{Q}_L \Phi_1 d_R + y_e \overline{L}_L \Phi_1 e_R \notag\\
        &+ y_\nu \overline{L}_L \tilde{\Phi}_2 N_R + \frac{1}{2} M_N \overline{N_R^c} N_R + \text{h.c.},
\end{align}
where $\tilde{\Phi}_i = i\sigma_2 \Phi^*_i$, and $y$ represents the Yukawa coupling matrices. Generation indices are suppressed. After electroweak symmetry breaking, the light neutrino mass matrix is then given by the seesaw formula with proper modification of VEV:
\begin{equation}\label{eq:light_nu_matrix}
    m_\nu = \frac{v_2^2}{2}y_\nu \mathcal{D}_M^{-1}y_\nu^T,
\end{equation}
where $\mathcal{D}_M = \text{diag}(M_1, M_2, M_3)$.

To explore the parameter space and ensure consistency with experimental data from neutrino oscillations, we employ the Casas-Ibarra~(CI) parametrization, which links low-energy neutrino data to high-energy seesaw parameters~\cite{Casas:2001sr}:
\begin{equation}\label{eq:yukawa_nu}
    y_\nu = \frac{\sqrt{2}}{v_2}U_\text{PMNS} \sqrt{\mathcal{D}_m} R^T\sqrt{\mathcal{D}_{M}},
\end{equation}
where $U_\text{PMNS}$ is the standard PMNS matrix with Majorana phases $\alpha_{21}$ and $\alpha_{31}$:
\begin{equation}
    U_\text{PMNS} = \hat{U}\cdot \text{diag}(1, e^{i\frac{\alpha_{21}}{2}}, e^{i\frac{\alpha_{31}}{2}}),
\end{equation}
and $R$ is a complex orthogonal matrix parameterized by three complex rotation angles~(or six real angles) $z_i = x_i + i y_i$~($i = 1, 2, 3$).

For RHNs, we consider the mass hierarchy $M_1 \ll T_\text{spha} < M_2 \sim M_3$. Such a pattern can arise from a softly-broken lepton flavor symmetry like $L_\alpha - L_\beta - L_\delta$~\cite{Lavoura:2000ci, Mohapatra:2001ns, Shaposhnikov:2006nn, Lindner:2010wr}, the Froggatt-Nielsen mechanism~\cite{Merle:2011yv}, or the split seesaw mechanism~\cite{Kusenko:2010ik}. In the case of a softly-broken flavor symmetry, the RHN mass pattern starts as $(0, M, M)$. Breaking the symmetry with a soft breaking term $S$ at a scale $\mathcal{O}(S)$ lifts the degeneracy to $(\mathcal{O}(S), M-\mathcal{O}(S), M+\mathcal{O}(S))$. For $M_2 \sim 10$~TeV with $\Delta M_{32}/M_2 \sim 10^{-3}$, this implies $M_1 \sim \mathcal{O}(\text{GeV}) \ll T_\text{spha}$.

This mass pattern is crucial: $N_1$ does not participate in thermal leptogenesis because when $N_1$ becomes thermally important~($T \sim M_1$), the temperature of the universe is already lower than $T_\text{spha}$, at which point all baryon asymmetry converted from lepton asymmetry is frozen. Additionally, in the decoupling limit of 2HDM, where the observed SM-like Higgs boson corresponds to the light neutral CP-even state~\cite{Haber:1994mt}, $m_H, m_A, m_{H^\pm} \gg m_h \sim 125$~GeV $\gg M_1$, there are no two-body tree-level decay channels for $N_1$. Three-body decays are, in principle, possible but are very suppressed via an off-shell heavy Higgs or the SM-like neutral Higgs. For example, if mediated by a heavy charged Higgs (e.g. $N_1 \to \nu\,\ell^+\ell^-$), the decay is highly suppressed by the off-shell propagator, and the decay width can be approximated as $\Gamma_{N_1} \sim \frac{|y_\nu|^4}{192\pi^3}\frac{M_1^5}{m_H^4}$. For TeV-scale heavy Higgs bosons and a sub-GeV $N_1$, this decay width is much smaller than the Hubble constant, ensuring that $N_1$ is effectively stable and contribute to the relic density. Similarly, three-body decays mediated by the SM-like neutral Higgs (e.g., $N_1 \to 3\nu$) are suppressed by extremely tiny couplings between the SM-like Higgs and the SM neutrinos. While the decays are highly suppressed, the production of $N_1$ is very efficient through 2$\to$2 scattering processes when the mediators are on-shell, ensuring that a sufficient relic abundance is produced. Thus, $N_1$ can serve as a dark matter candidate. Our model, therefore, has the potential to simultaneously address the BAU (via the mildly degenerate $N_2$ and $N_3$) and the observed dark matter relic density (through the stable $N_1$) without introducing additional fields or mechanisms.

\section{Dynamics of Leptogenesis}\label{sec:dyna_lepto}

The role of lepton flavor becomes critical in leptogenesis when the charged lepton interactions with the thermal bath enter equilibrium. At very high temperatures, these interactions are negligible compared with the universe’s expansion rate $H(T)$, and all lepton flavors remain coherent. As the universe cools, the Yukawa interactions of charged leptons come into equilibrium one by one. By $T \sim 10^{12}\text{ GeV}$, the $\tau$ Yukawa coupling becomes strong enough to bring the $\tau$ flavor into thermal equilibrium. At $T \sim 10^9\text{ GeV}$, the $\mu$ Yukawa interaction also comes into equilibrium. All three lepton flavors are then distinguished~\cite{Nardi:2005hs}. 

As in standard resonant leptogenesis, the mild-degeneracy of the RHNs $N_2$ and $N_3$ leads to a resonant enhancement of the CP asymmetry from the self-energy diagram contributions. Also, for TeV-scale leptogenesis concerned, all three flavors are fully resolved. In this flavored regime, the CP asymmetry generated by the decay of $N_i$ into a specific lepton flavor $\alpha = e, \mu, \tau$ must be computed separately. The resonant flavored CP asymmetry $\epsilon_{i\alpha}$ from the decay of $N_i$ into $L_\alpha$ is given by~\cite{DeSimone:2007edo}
\begin{equation}\label{eq:cp_res}
    \begin{split}
        -\epsilon_{i\alpha} &= \sum_{j \neq i}  (f^\text{mix}_{ij} + f^\text{osc}_{ij})\\
        &\times\frac{\text{Im}\left[ (y_\nu^\dagger)_{i\alpha} (y_\nu)_{\alpha j} (y_\nu^\dagger y_\nu)_{ij} + \frac{M_i}{M_j} (y_\nu^\dagger)_{i\alpha} (y_\nu)_{\alpha j} (y_\nu^\dagger y_\nu)_{ji} \right]}{(y_\nu^\dagger y_\nu)_{ii}(y_\nu^\dagger y_\nu)_{jj}},\\
        f^\text{mix}_{ij} &= \frac{(M_i^2 - M_j^2)M_i\Gamma_j}{(M_i^2 - M_j^2)^2 + M_i^2\Gamma_j^2},\\
        f^\text{osc}_{ij} &= \frac{(M_i^2 - M_j^2)M_i\Gamma_j}{(M_i^2 - M_j^2)^2 + (M_i\Gamma_i + M_j\Gamma_j)^2\frac{\text{det}\left[\text{Re}(y_\nu^\dagger y_\nu)\right]}{(y_\nu^\dagger y_\nu)_{ii}(y_\nu^\dagger y_\nu)_{jj}}},
    \end{split}
\end{equation}
where $\Gamma_i$ is the total decay width of $N_i$.

The full dynamics of lepton asymmetry in the flavored regime are governed by the density matrix formalism, which tracks flavor-specific asymmetries through diagonal elements and coherence effects through off-diagonal elements. However, for TeV-scale leptogenesis, coherence between different lepton flavors is entirely destroyed by thermal interactions damping the off-diagonal elements of the density matrix. Consequently, the differential equations for the density matrix reduce to a set of flavor-diagonal Boltzmann equations for each lepton flavor asymmetry. This approximation is well-valid for $M_N \ll 10^9\text{ GeV}$~\cite{Blanchet:2011xq, Moffat:2018wke}.

Assuming $N_2$ and $N_3$ decay predominantly into nearly orthogonal lepton flavors, e.g., $N_2 \to L_\alpha$ and $N_3 \to L_\beta$~($\alpha \neq \beta$), the interaction rates of $N_3$ with $L_\alpha$ are negligible. This ensures that the lepton number excess in the $\alpha$ flavor generated by $N_2$ decays is not erased by $N_3$-mediated washout~\cite{Pilaftsis:2004xx, Pilaftsis:2003gt}. This assumption requires
\begin{equation}\label{eq:flavor_assump}
    \epsilon_{2\alpha} \gg \epsilon_{3\alpha}, \quad p_{3\alpha}  = \frac{|(y_\nu)_{\alpha3}|^2}{(y_{\nu}^\dagger y_\nu)_{33}}  \ll 1,
\end{equation}
where $p_{i\alpha}$ encodes the flavor component $\alpha$ in $N_i$. Under this condition, we can focus solely on Boltzmann equations for the $N_2$ system, which are
\begin{equation}\label{eq:BE}
    \begin{split}
        \frac{dn_{N_2}}{dz} &= - D_2 (n_{N_2} - n_{N_2}^{\text{eq}}),\\
        \frac{dn_{\alpha}}{dz} &=  \epsilon_{2\alpha} D_2 (n_{N_2} - n_{N_2}^{\text{eq}}) - p_{2\alpha}W_2 n_{\alpha},
    \end{split}
\end{equation}
where $n_{N_2}$ is the number density of $N_2$, $n_{N_2}^\text{eq}$ is its equilibrium value, $n_\alpha$ is the lepton asymmetry in $\alpha$ flavor, and $z = M_2 / T$. The decay~($D_i$) and washout~($W_i$) rates of $N_i$ are~\cite{Buchmuller:2004nz, Davidson:2008bu}
\begin{align}
    D_i(z) &= \sum_\alpha D_{i\alpha} = \sum_\alpha K_{i\alpha} x_iz\frac{\mathcal{K}_1(z)}{\mathcal{K}_2(z)},\\
    W_i(z) &= \frac{1}{4}K_i\sqrt{x}\mathcal{K}_1(z)z^3,
\end{align}
where $\mathcal{K}_1$ and $\mathcal{K}_2$ are modified Bessel functions of the second kind, $x_i = M_i^2/M_2^2$, and
\begin{equation}\label{eq:K}
    K_i = \sum_\alpha K_{i\alpha} \equiv \frac{\sum_\alpha\Gamma_{i\alpha}}{H(T = M_2)} = \frac{\sum_{\alpha}\tilde{m}_{i\alpha}}{m_*}
\end{equation}
tells whether the decay of $N_i$ is in equilibrium at $T = M_2$. It is related to the effective neutrino mass $\tilde{m}_{i\alpha}$ and the equilibrium neutrino mass $m_*$, given by
\begin{equation}\label{eq:eff_equi_m_nu}
    \tilde{m}_{i\alpha} \equiv \frac{(y_\nu)^\dagger_{i\alpha} (y_\nu)_{\alpha i}v_2^2}{M_i}, \quad
    m_* \equiv \sqrt{\frac{8\pi^3 g_*}{90}} \frac{8\pi v_2^2}{M_\text{Pl}},
\end{equation}
where $g_*$ is the total number of relativistic degrees of freedom, and $M_\text{Pl}$ is the Planck mass.

For large $\tan\beta$, $v_2 \ll v_\text{SM}$ significantly reduces $m_*$, favoring the strong washout regime~($K_2 = \tilde{m}_2 / m_* > 1$) \footnote{The effective neutrino mass $\tilde{m}_2$ does not depend on $v_2$, as it cancels with the one in the Yukawa matrix, which can be seen from Eq.~\eqref{eq:yukawa_nu}.}. Unlike in the weak washout regime, $\Delta L = 1$ scattering effects are subdominant in the strong washout regime, and affect the final lepton asymmetry only at the $\mathcal{O}(10)\%$ level, justifying why we ignore such terms in Eq.~\eqref{eq:BE} to simplify analysis~\cite{Blanchet:2006be, Frossard:2013bra}. Spectator and thermal corrections are also omitted as they do not qualitatively alter leptogenesis dynamics at low scales.

However, as in standard $\nu$2HDM models, $\Delta L = 2$ scattering processes become increasingly relevant at large $\tan\beta$, due to the enhancement on the Yukawa couplings of neutrinos~\cite{Clarke:2015hta}. These processes can significantly wash out the lepton asymmetry generated. They involve $\Phi_2 \ell \leftrightarrow \bar{\Phi}_2\bar{\ell}$ and $\Phi_2\Phi_2 \leftrightarrow \ell\ell$, with rates proportional to $\Tr[(y_\nu y_\nu^T)(y_\nu y_\nu^T)^\dagger] \sim M_2^2 \bar{m}^2 / v_2^4$, where $\bar{m}^2 = \sum_i m_i^2 \sim 0.05\text{ eV}$ represents the light neutrino mass scale. At temperatures $T \lesssim M_2 / 3$, the thermally averaged $\Delta L = 2$ scattering rate is approximately~\cite{Buchmuller:2004nz, Clarke:2015hta}
\begin{equation}
     \frac{\Gamma_{\Delta L = 2}}{H} \approx\frac{T}{2.2\times 10 ^{13}\text{ GeV}}\left(\frac{246 \text{ GeV}}{v_2}\right)^4\left(\frac{\bar{m}}{0.05 \text{ eV}}\right)^2.
\end{equation}
This scales as $1/v_2^4$, making $\Delta L = 2$ processes particularly significant in the small $v_2$~(large $\tan\beta$) regime. In this region, the washout effect can become so strong that it potentially erases most of the lepton asymmetry. To ensure successful leptogenesis, we exclude parameters in strong $\Delta L = 2$ washout regime from our analysis.

\begin{figure}[htb!]
    \centering
    \includegraphics[width=0.45\textwidth]{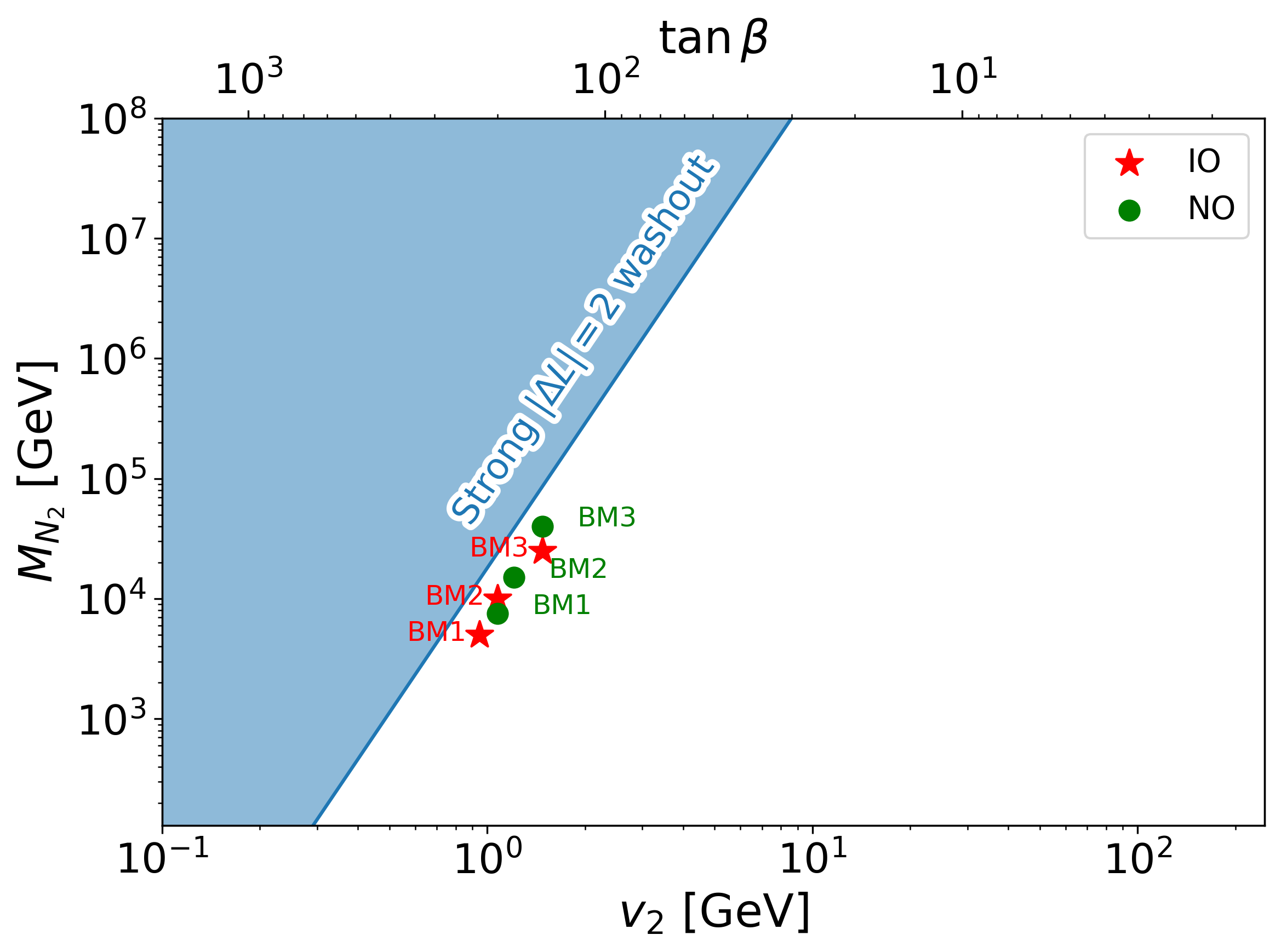}
    \caption{Bounds on $M_2$ by $\Delta L = 2$ washout rate as a function of $v_2$. Locations of BM points for successful leptogenesis from Table \ref{tab:BM_points_simple} are also shown.}
    
    \label{fig:deltaL2}
\end{figure}

Fig.~\ref{fig:deltaL2} illustrates the bounds on $M_2$ imposed by $\Delta L = 2$ washout as a function of $v_2$. Notably, while the Type-I 2HDM does not impose an upper limit on $\tan\beta$, requiring leptogenesis to occur before sphaleron freeze-out while avoiding strong $\Delta L = 2$ washout places a lower bound on $v_2 \gtrsim 0.3\text{ GeV}$~($\tan\beta \lesssim 600$).

Finally, the generated lepton asymmetry is partially converted into baryon asymmetry via $B - L$ conserving electroweak sphalerons, which remain efficient until $T_\text{spha}$. The baryon-to-photon ratio is related to the $B-L$ asymmetry by~\cite{Buchmuller:2004nz}
\begin{equation}\label{eq:spha_convert}
    \eta_B \equiv \frac{n_B}{n_{\gamma}^{\text{rec}}} = 0.013 n_{B-L},
\end{equation}
where $n_{B-L} = \sum_{\alpha=e,\mu,\tau} n_{\alpha}$ is the total $B-L$ asymmetry and $n_{\gamma}^{\text{rec}}$ is the photon number density at recombination.

\section{Numerical Analysis}\label{sec:analysis}

To evaluate the viability of our model, we performed a comprehensive numerical scan of the parameter space relevant to leptogenesis and neutrino oscillation data to identify the preferred regions for the mixing angles and phases.

For neutrino oscillation inputs, we used NuFit-6.0 without Super-Kamiokande's atmospheric data~\cite{Esteban:2024eli}. The light neutrino mass-squared differences $\Delta m_{21}^2$ and $\Delta m_{3\ell}^2$, where $\ell=1$ for normal ordering~(NO) and $\ell=2$ for inverted ordering~(IO), were fixed at their best-fit values. The Dirac phase $\delta_\text{CP}$ was scanned over its $3\sigma$ range, while the PMNS mixing angles $\theta_{12}$, $\theta_{23}$, and $\theta_{13}$ were scanned within their best-fit $\pm 1\sigma$ intervals. A uniform random scan was also performed over the following parameters:
\begin{itemize}
    \item $x_i, y_i$: $-180^\circ - 180^\circ$,
    \item $\alpha_{21}, \alpha_{31}$: $0^\circ - 720^\circ$,
    \item $\log_{10}(M_1 \, [\text{GeV}])$: $-7.0 - 2.0$,
    \item $\log_{10}(m_{1(3)} \, [\text{eV}])$: $-10.0- (-1.0)$,
\end{itemize}
while $M_2=10$ TeV, $\Delta M_{32}/M_2 = 0.1\%$, and $\tan\beta=225$ were kept fixed. The Yukawa matrix elements were then reconstructed from Eq.~\eqref{eq:yukawa_nu}, and the Boltzmann equations for the lepton asymmetry were solved numerically for each scan point. The resultant lepton asymmetry was converted into the final baryon asymmetry via Eq.~\eqref{eq:spha_convert} and compared with the present-day observed value $\eta_B \approx 6\times 10^{-10}$. 

\begin{figure*}[htb!]
    \centering
    \begin{subfigure}{0.45\textwidth}
        \centering
        \includegraphics[width=\textwidth]{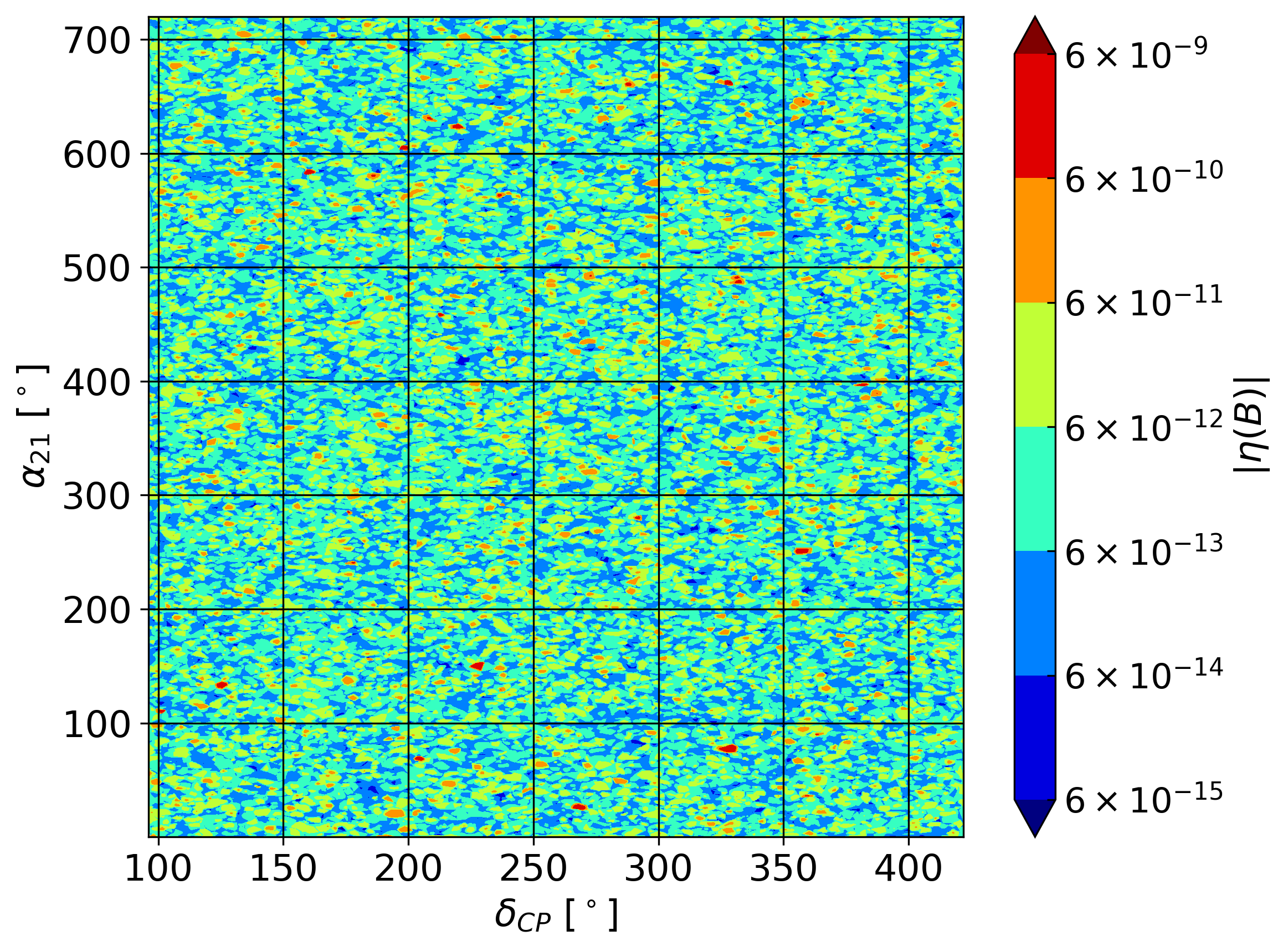}
        \caption{NO}
    \end{subfigure}
    \begin{subfigure}{0.45\textwidth}
        \centering
        \includegraphics[width=\textwidth]{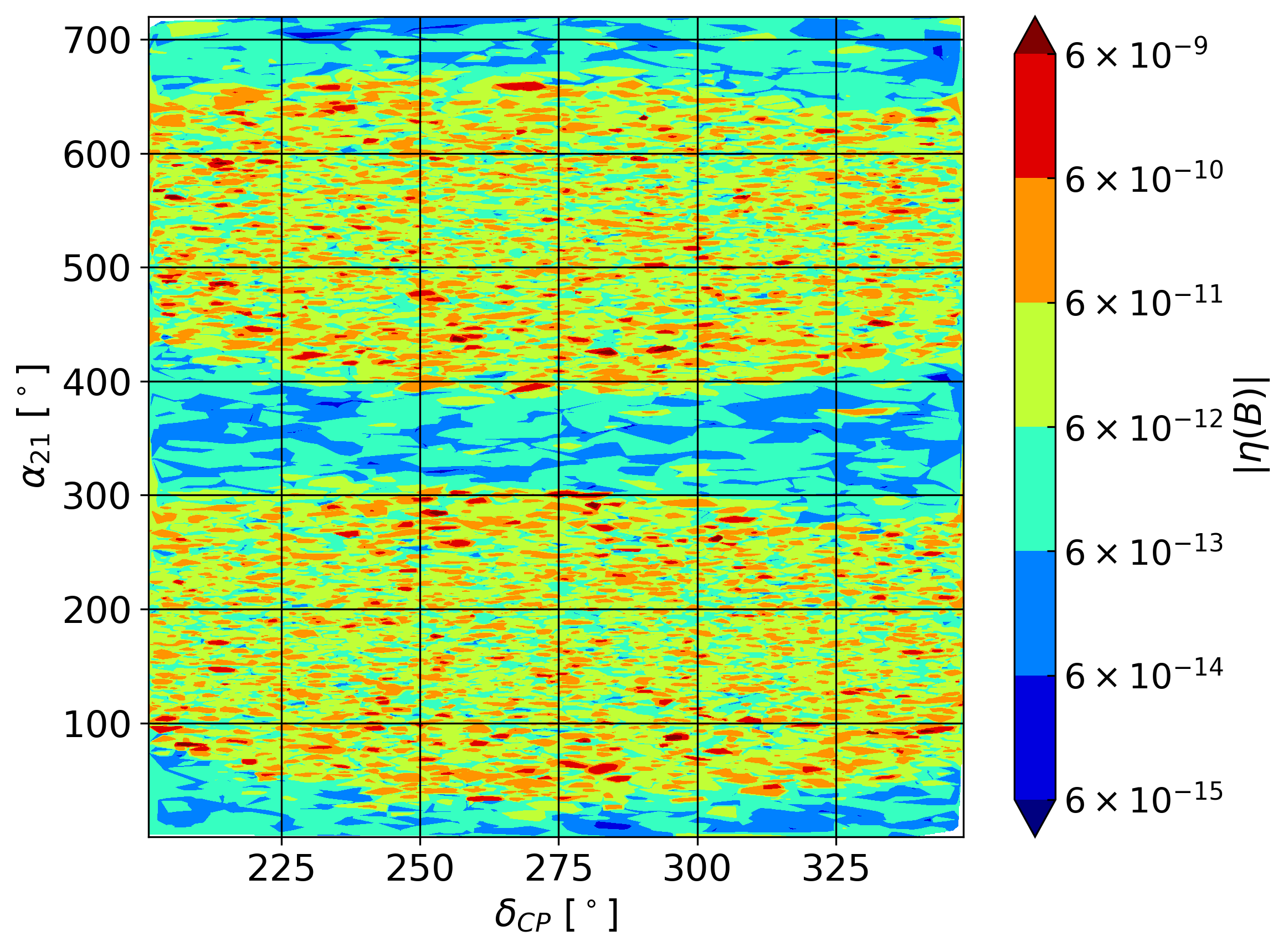}
        \caption{IO}
    \end{subfigure}
    \caption{BAU dependence as a contour on $\delta_\text{CP}$ and $\alpha_{21}$ plane. The color corresponds to the magnitude of $\eta_B$ survived. The boundary of red and orange corresponds to the observed value.}
    \label{fig:low_CP}
\end{figure*}

The results are displayed as color contour maps to reveal patterns and correlations among the scanned parameters, particularly the mixing angles and phases. For instance, Fig.~\ref{fig:low_CP} shows the contour on the plane of $\alpha_{21}$ versus $\delta_\text{CP}$ under the NO and IO spectrum, respectively. The color represents the magnitude of the final baryon asymmetry $|\eta_B|$. The observed $\eta_B$ corresponds to the boundary of red and orange. For the NO spectrum, the generated BAU does not depend on either Dirac or Majorana CP phases. For the IO spectrum, a clear preference for $\alpha_{21} \in [45^\circ, 315^\circ] \cup [405^\circ, 675^\circ]$ is evident while $\delta_\text{CP}$ shows no strong preference. In both spectrum, $\alpha_{31}$ was also found to be irrelevant to the baryon asymmetry.

\begin{figure*}[htb!]
    \centering
    \begin{subfigure}{0.45\textwidth}
        \centering
        \includegraphics[width=\textwidth]{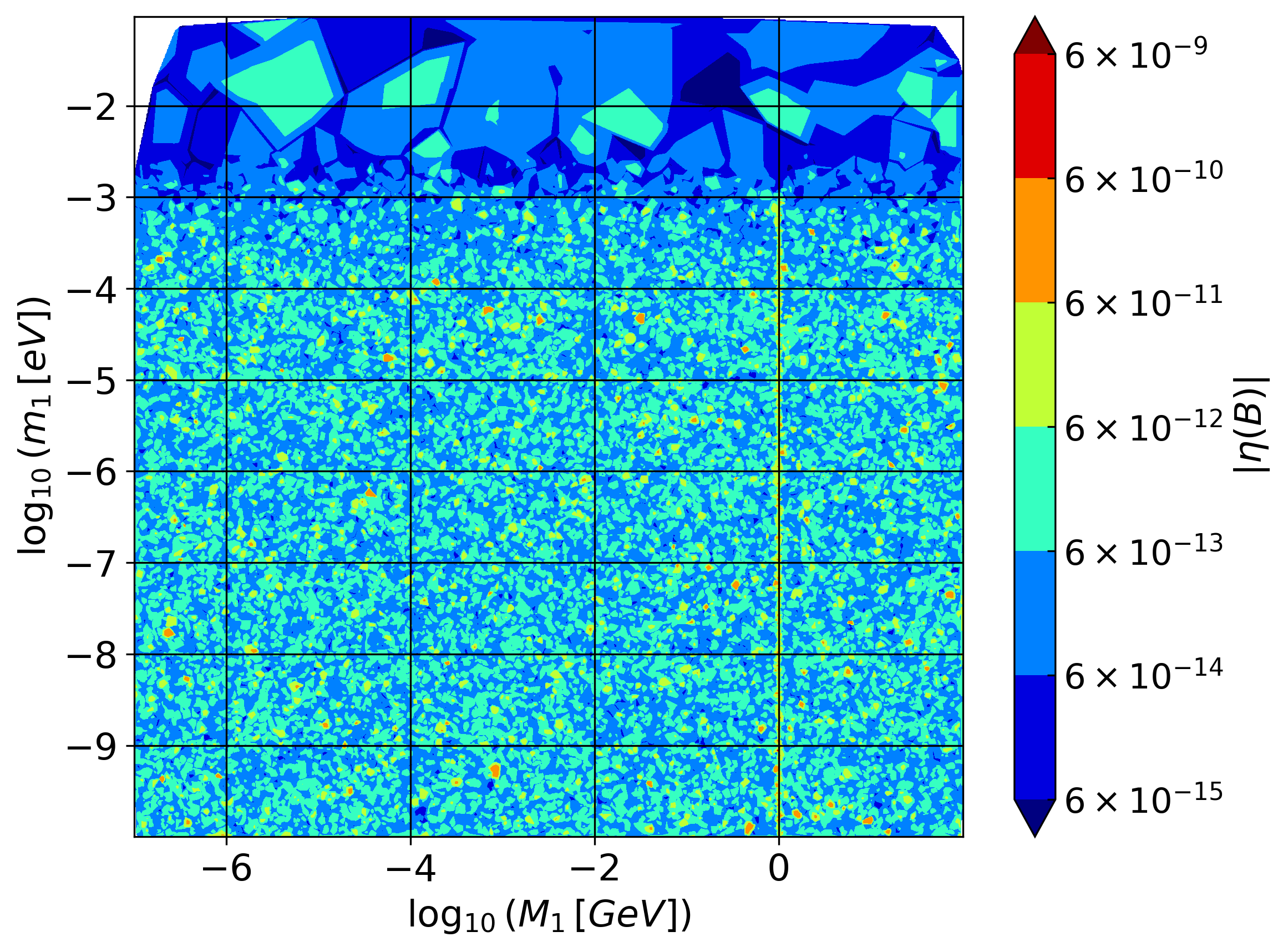}
        \caption{NO}
    \end{subfigure}
    \begin{subfigure}{0.45\textwidth}
        \centering
        \includegraphics[width=\textwidth]{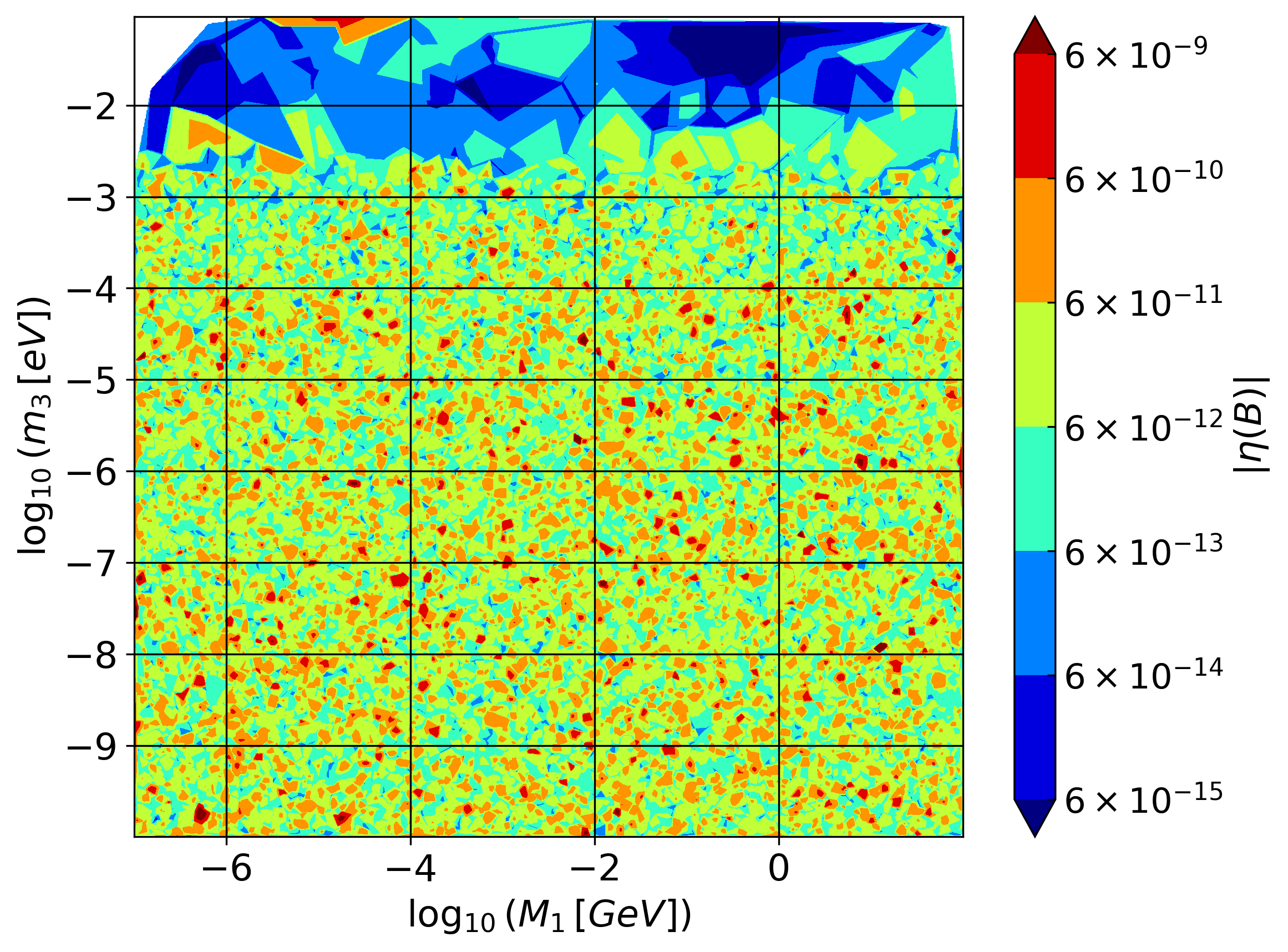}
        \caption{IO}
    \end{subfigure}
    \caption{BAU dependence as a contour on the mass of the lightest LH neutrino $m_{1(3)}$ and the mass of the lightest RH neutrino $M_1$.}
    \label{fig:low_scale_masses}
\end{figure*}

Fig.~\ref{fig:low_scale_masses} shows the contour scan of the lightest left-handed neutrino mass $m_{1(3)}$ and right-handed neutrino mass $M_{1}$. Our scans show that $M_1$ has no influence on the baryon asymmetry. However, the lightest light neutrino mass $m_{1(3)}$ exhibits a threshold for generating sufficient baryon asymmetry. For $\tan\beta=225$, $m_{1(3)} \lesssim 10^{-3}$~eV is required. This behavior arises because a smaller $v_2$ reduces the equilibrium neutrino mass $m_*$, necessitating a reduction in the effective neutrino mass $\tilde{m}_{2\alpha}$ to counteract enhanced washout. Since $\tilde{m}_{2\alpha}$ does not depend on $v_2$, the suppression of the lightest neutrino mass becomes essential to partially balance the washout factor $K$ in Eq.~\eqref{eq:K} for fixed $M_2$.

\begin{figure*}[htb!]
    \centering
    \begin{subfigure}{\textwidth}
        \centering
        \includegraphics[width=\textwidth]{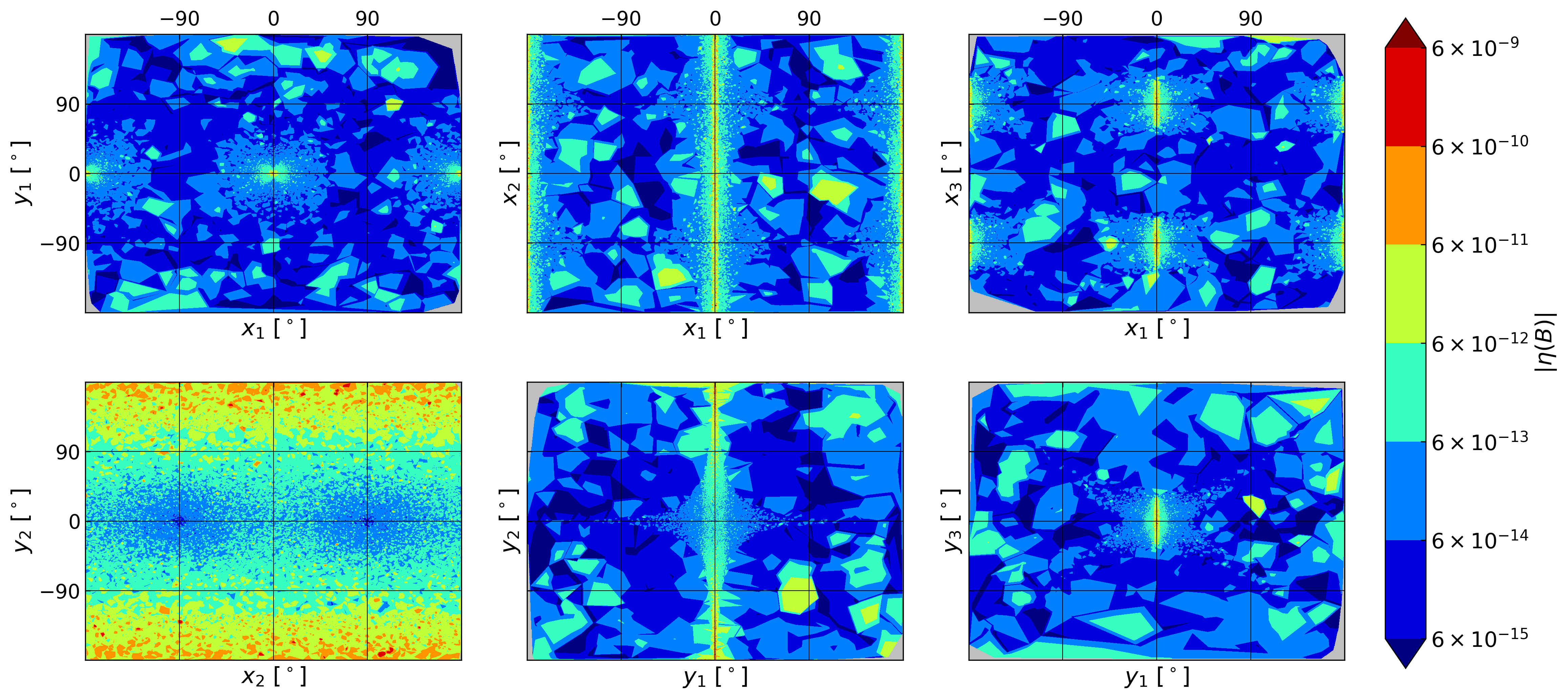}
        \caption{$\tan\beta=225$}
    \end{subfigure} \\
    \begin{subfigure}{\textwidth}
        \centering
        \includegraphics[width=\textwidth]{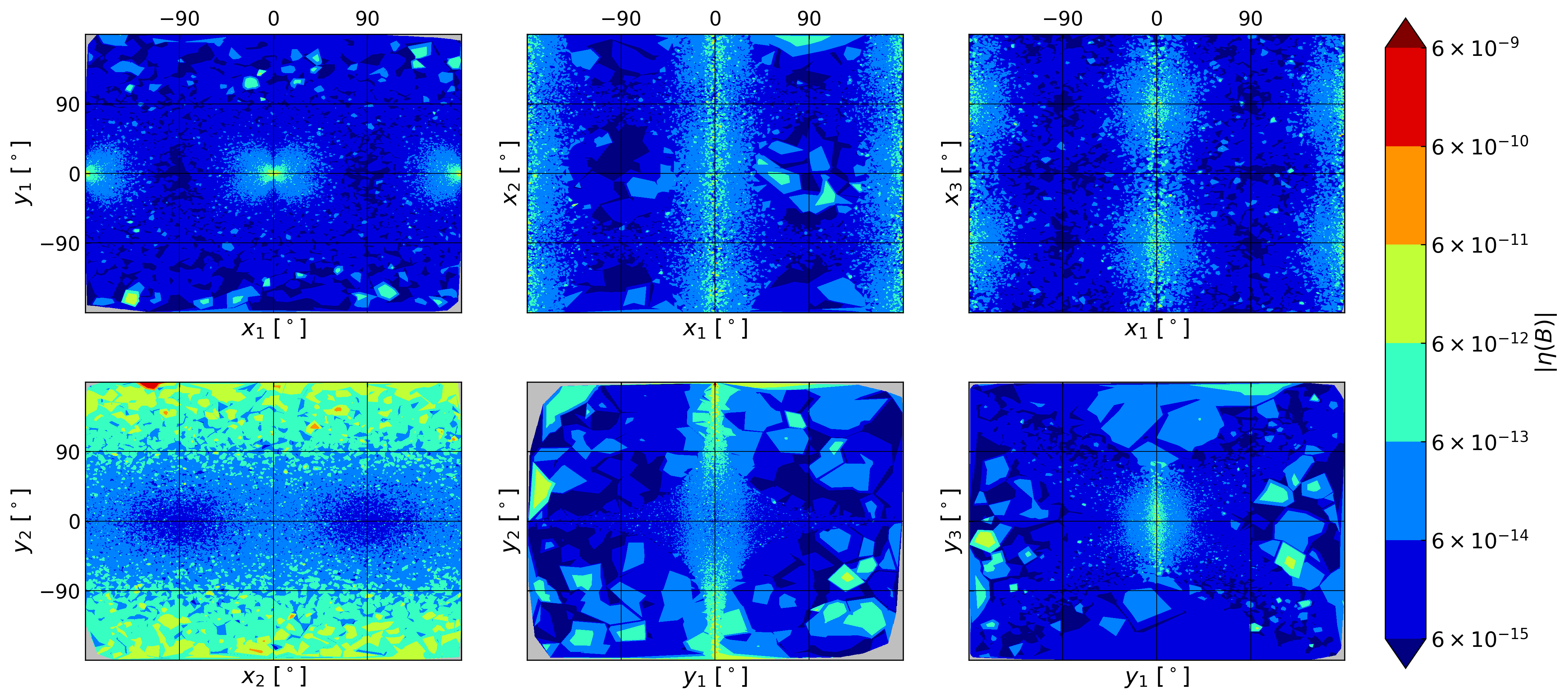}
        \caption{$\tan\beta=100$}
    \end{subfigure}
    \caption{Scan contours for NO spectrum with $\tan\beta=225,\ 100$.}
    \label{fig:scan_NO}
\end{figure*}
\begin{figure*}[htb!]
    \centering
    \begin{subfigure}{\textwidth}
        \centering
        \includegraphics[width=\textwidth]{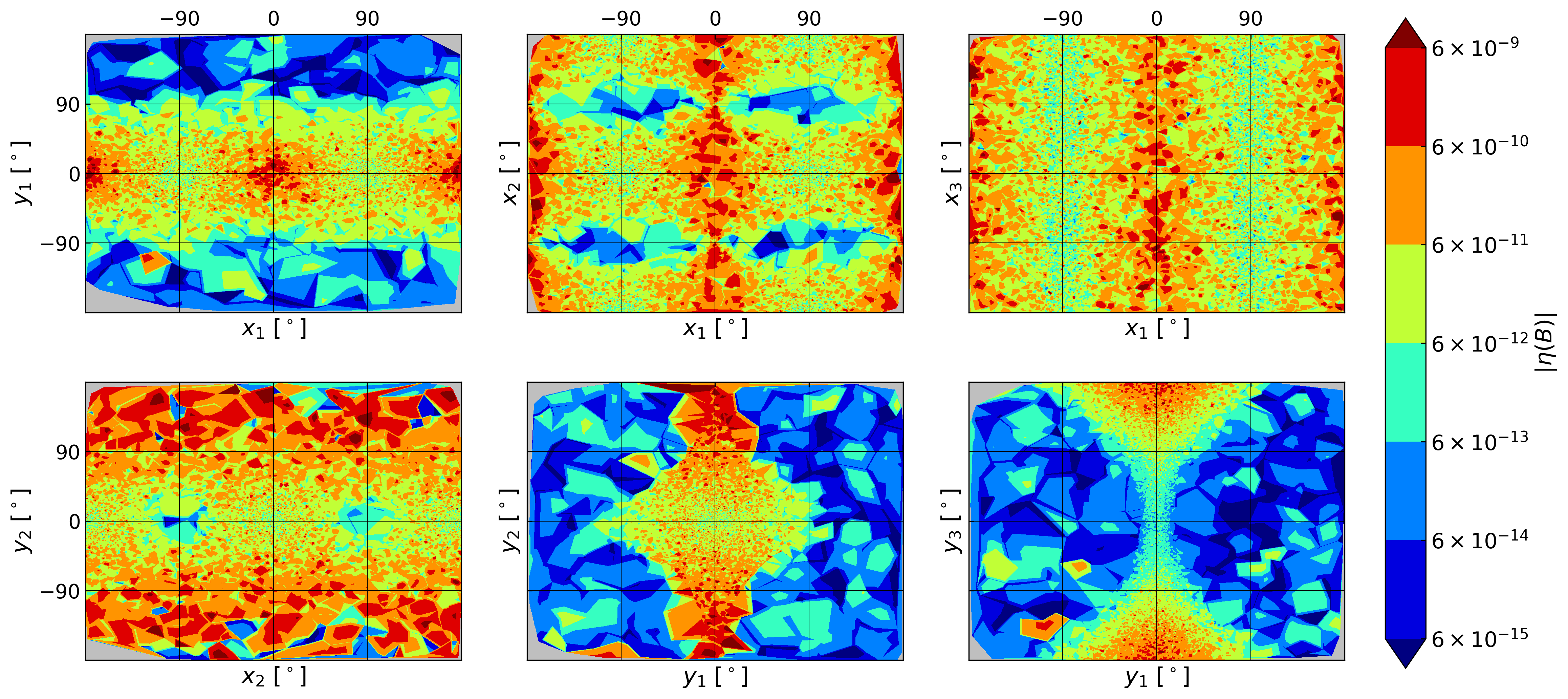}
        \caption{$\tan\beta=225$}
    \end{subfigure} \\
    \begin{subfigure}{\textwidth}
        \centering
        \includegraphics[width=\textwidth]{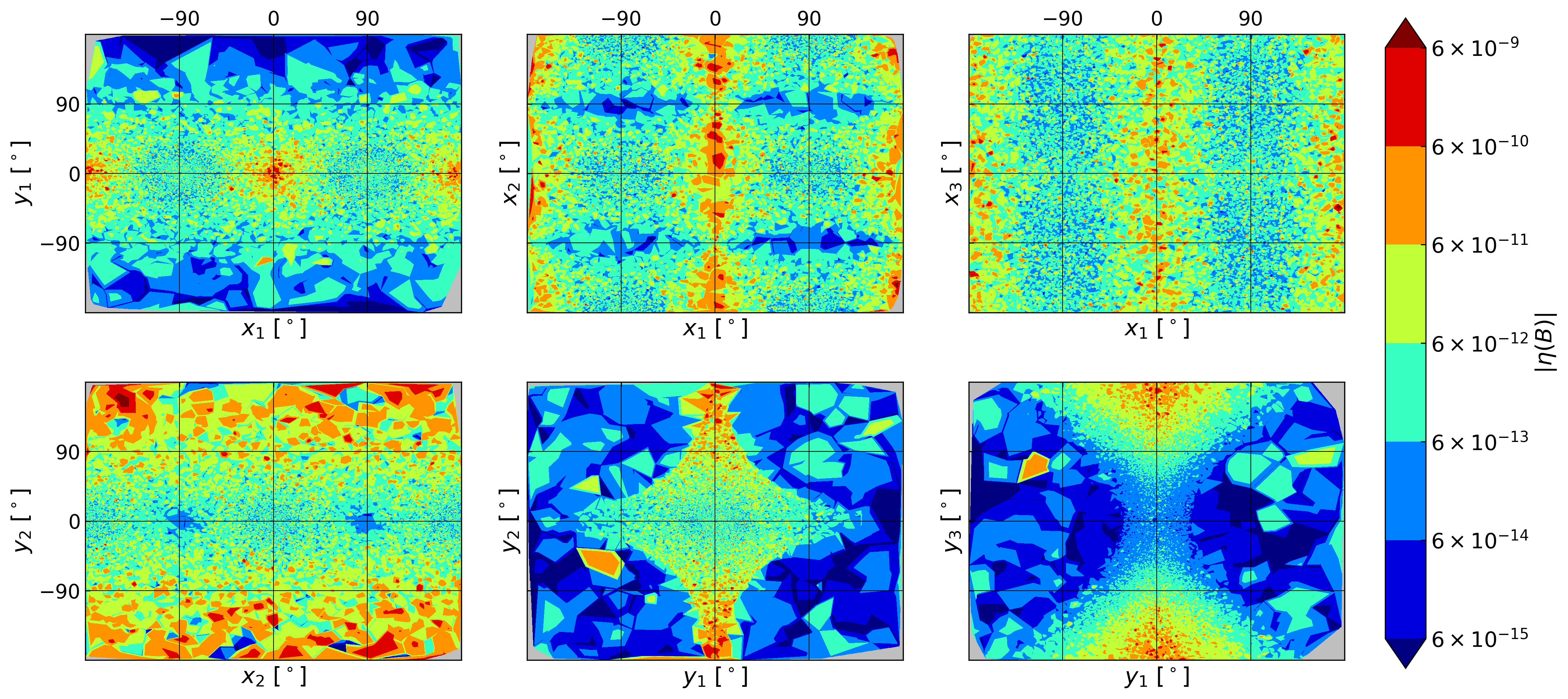}
        \caption{$\tan\beta=100$}
    \end{subfigure}
    \caption{Scan contours for IO spectrum with $\tan\beta=225,\ 100$.}
    \label{fig:scan_IO}
\end{figure*}

The reason we fixed the values of $M_2$, $\Delta M_{32}/M_2$, and $\tan\beta$ for the above scan is these parameters influence the leptogenesis dynamics in a flavor-blind manner. Varying them only shifts the boundaries of amount of BAU~(quantified by color layers) on the contour plots without altering the internal correlations and patterns among the mixing angles and phases.

A detailed scan contours for the Casas-Ibarra mixing angles $x_i$ and $y_i$ for both NO and IO neutrino spectra is also performed. Fig.~\ref{fig:scan_NO} shows the scan results for the NO spectrum. Panel (a) depicts the contours for $\tan\beta = 225$, while panel (b) shows the results for $\tan\beta = 100$ to illustrate the impact of varying $\tan\beta$. As mentioned above, varying $\tan\beta$ only shifts the boundaries of the amount of BAU on the contour plots without altering the internal correlations and patterns among the mixing angles and phases. In general, for NO,
\begin{itemize}
    \item $|x_1|$ tends to align closely with $0^\circ$ or $180^\circ$, while $|y_1|$ clusters around $0^\circ$. Moreover, $x_1$ and $y_1$ either align or anti-align with each other.
    \item $|y_2|$ prefers larger values.
    \item $|x_3|$ clusters near $90^\circ$, 
    \item $|y_3|$ prefers small values.
    \item $x_2$, $\delta_\text{CP}$, $\alpha_{21}$, and $\alpha_{31}$ do not exhibit a preference for leptogenesis.
\end{itemize}
Fig.~\ref{fig:scan_IO} presents the scan results for the IO spectrum. We observed that
\begin{itemize}
    \item $|x_1|$ and $|y_1|$ tend to cluster near $0^\circ$.
    \item $|y_2|$ shows an inverse correlation with $|y_1|$ and prefers large values.
    \item $|x_2|$ avoids $90^\circ$.
    \item $|y_3|$ favors large values and exhibits an angular shift relative to $|y_1|$.
    \item Unlike NO, $\alpha_{21}$ shows a clear preference for $[45^\circ, 315^\circ] \cup [405^\circ, 675^\circ]$, while $x_3$, $\delta_\text{CP}$, and $\alpha_{31}$ remain irrelevant for leptogenesis.
\end{itemize}

These contours reveal highly structured patterns and correlations among the CI mixing angles for successful BAU. Notably, these patterns actually tend to select specific flavors responsible for leptogenesis, depending on the neutrino mass ordering. For NO, successful leptogenesis has a strong taste of the $e$ flavor for generating the excess, while for IO, it has a distaste of the $e$ flavor and instead nearly equally favors $\mu$ or $\tau$ flavors. These distinct patterns may suggest a potential underlying flavor symmetry that constrains the mixing angles. We leave the possibility that the same symmetry could also govern the RHN mass patterns. Further investigation into this flavor symmetry could provide valuable insights into the interplay between mixing angles and mass patterns in the RHN sector. The contours also suggest the IO case allows a richer parameter space than the NO case.

Based on the scan results, we identified targeted benchmark points (BMs) to demonstrate the viability of leptogenesis within our model. These points satisfy neutrino oscillation experimental constraints, generate sufficient BAU, and lie outside the strong $\Delta L = 2$ washout region shown in Fig.~\ref{fig:deltaL2}. The BMs were chosen to reflect distinct levels of $M_2$, mass degeneracy levels, and $\tan\beta$: either {\it a}) small $M_2$ with high degeneracy (still mild in general terms), {\it b}) large $M_2$ with low degeneracy, or {\it c}) a trade-off between these two directions. The BMs are summarized in Table~\ref{tab:BM_points_simple}, while a complete set of parameters for these BMs can be found in the appendix~\ref{app:bm_points}. 

\begin{figure*}[htb!]
    \centering
    \includegraphics[height=0.25\textheight, width=\textwidth]{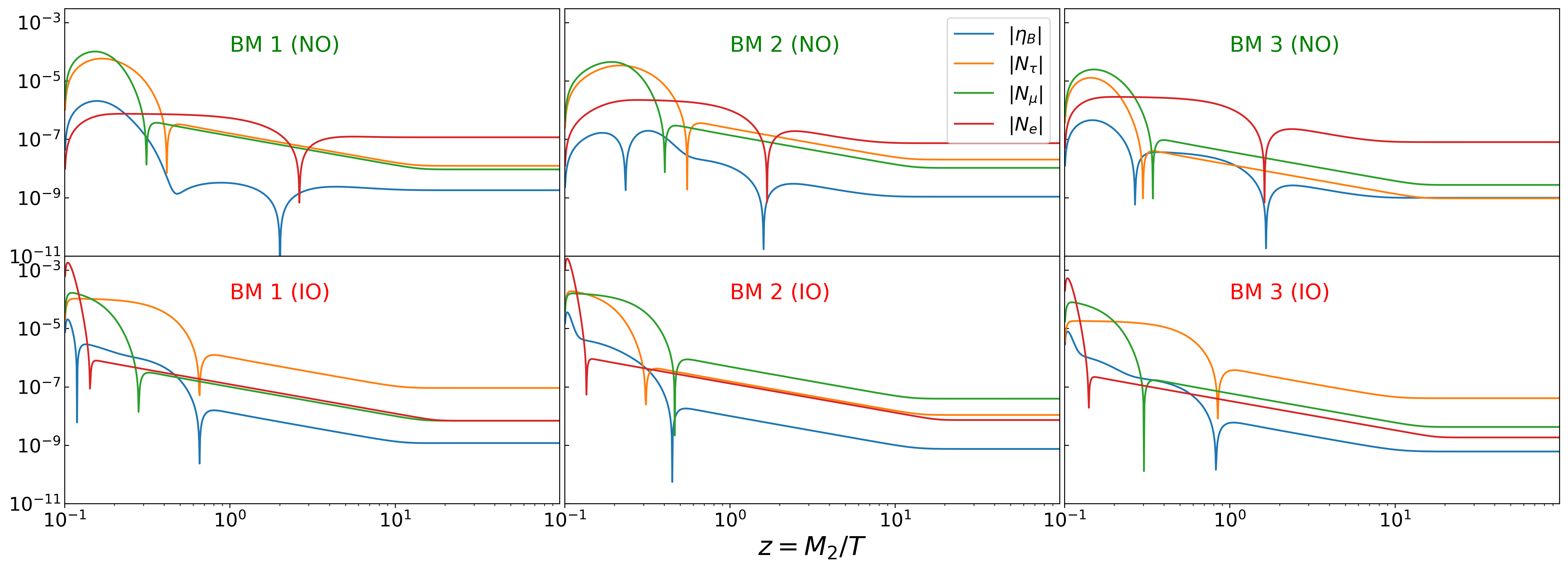}
    \caption{Number densities evolution as a function of $z = M_2/T$.}    
    \label{fig:BM_evo}
\end{figure*}
\begin{table}[htb!]
    \centering
    \begin{tabular}{|c|c c c|}
        \hline
        \textbf{NO} & \textbf{BM 1} & \textbf{BM 2} & \textbf{BM 3}\\
        \hline
        \hline
        $M_2$ [TeV] & 7.5 & 15 & 40 \\
        \hline
        $\Delta M_{32}/ M_2$ & 0.1\% & 0.5\% & 1\% \\
        \hline
        $\tan\beta$ & 200 & 180 & 150 \\
        \hline
        \hline
        \textbf{IO} & \textbf{BM 1} & \textbf{BM 2} & \textbf{BM 3}\\
        \hline
        \hline
        $M_2$ [TeV] & 5 & 10 & 25 \\
        \hline
        $\Delta M_{32}/ M_2$ & 1\% & 4\% & 10\% \\
        \hline
        $\tan\beta$ & 225 & 200 & 150 \\
        \hline
    \end{tabular}
    \caption{Benchmark points for normal and inverted neutrino ordering. $M_1 = 1$ GeV is fixed for all BM points.}
    \label{tab:BM_points_simple}
\end{table}

The evolution of the number densities for these BMs is shown in Fig.~\ref{fig:BM_evo}. These evolutions are characterized by the development of two out of three lepton excesses that build up early and are subsequently washed out. Meanwhile, the third flavor excess remains resilient, ultimately surviving and contributing to the final baryon asymmetry observed today, aligning with the flavor assumption in Sec.~\ref{sec:dyna_lepto}. 

These results establish the compatibility of resonant leptogenesis down to TeV scales with only a very mild mass degeneracy $\mathcal{O}(1\%-0.1\%)$ needed. For NO, this can be achieved with $M_2=7.5$~TeV and $\Delta M_{32}/M_2\sim 0.1\%$, while for IO, $M_2=5$~TeV and $\Delta M_{32}/M_2\sim 1\%$.

\section{Conclusion}

In this study, we have demonstrated that TeV-scale leptogenesis can be achieved with only a very mild mass degeneracy between the RHNs $N_2$ and $N_3$. This degeneracy can naturally arise, for instance, from a flavor symmetry whose soft breaking results in the required mass pattern. Such a flavor symmetry might also explain the structured patterns observed in the scan of Casas-Ibarra rotation angles. A UV-complete realization of this scenario is left for future investigation. Importantly, in such a scenario, the mass of the lightest right-handed neutrino, $N_1$, is naturally below both the sphaleron freeze-out temperature $T_\text{spha}$ and the SM-like Higgs boson mass $m_h$.

Traditional leptogenesis models often face challenges in simultaneously addressing the BAU and the dark matter relic density, as they typically rely on the decay of the lightest RHN, $N_1$, or risk $N_1$ washing out the asymmetry generated by $N_2$ and $N_3$. Our model circumvents this issue by placing $N_1$ below $T_\text{spha}$, rendering it inactive during leptogenesis. Furthermore, since $N_1$ is lighter than the SM-like Higgs boson, it lacks 2-body tree-level decay channels while 3-body tree-level is very suppressed, making it a viable DM candidate. Thus, both BAU and potentially the DM relic density can be accounted for.

Unlike many other TeV-scale resonant leptogenesis models, which decouple the heaviest RHN from dynamics, our model keeps all RHNs at or below the TeV scale. Thus, all model parameters are, in principle, testable in near-future experiments. The three RHNs can be produced and probed directly in future collider in principle~\cite{Haba:2011nb, Huitu:2017vye, Chekkal:2017eka}. The MeV-GeV scale $N_1$ particularly offers lucrative experimental prospects. Depending on its mixing with light neutrinos, it could be produced in fixed-target experiments such as SHiP~\cite{SHiP:2015vad} and NA62~\cite{NA62:2017rwk}, or in rare meson decays at collider experiments, which it may manifest through displaced vertices or exotic decays of mesons. It could also be searched at FASER~\cite{FASER:2019dxq}, MATHUSLA~\cite{MATHUSLA:2019qpy}, DUNE~\cite{DUNE:2020ypp, Dutta:2024hqq, Dutta:2024kuj}, and FPF~\cite{Anchordoqui:2021ghd, Feng:2022inv}or direct DM detection experiments like SENSEI, which utilize recoils resulting from dark matter-electron interactions, are particularly sensitive to DM candidates in the MeV–GeV mass range~\cite{SENSEI:2023zdf}. Depending on the specific flavor assignments of each RHN, the model can also predict lepton flavor-violating (LFV) processes such as $\mu \to e\gamma$, $\mu \to 3e$, and $\mu \to e$ conversion in nuclei~\cite{Pilaftsis:2005rv, Chekkal:2017eka}, or LFV processes involving $\tau$. In particular, there can be sizable contributions to neutrinoless double beta decay~($0\nu\beta\beta$). The effective neutrino mass involved in $0\nu\beta\beta$ is related to the $(1,1)$ entry of the light neutrino mass matrix~\cite{Pilaftsis:2005rv}. As a concrete example, for BM 1 (IO) in Table~\ref{tab:BM_points_simple}, we calculated that
\begin{equation}
    |\langle m_{\beta\beta}\rangle| = |(m_\nu)_{11}| \approx 0.024 \, \text{eV},
\end{equation}
which is within the sensitivity of upcoming ton-scale experiments~\cite{Gomez-Cadenas:2019sfa, LEGEND:2017cdu, nEXO:2017nam, Han:2017fol, Armengaud:2019loe, Paton:2019kgy}.

Finally, the simplicity of our approach is noteworthy. By making minimal and well-motivated extensions, we provide a unified framework that simultaneously addresses the BAU, neutrino masses, and DM. This is both theoretically appealing and experimentally testable.

\begin{acknowledgments}
We would like to thank K.S. Babu for helpful discussions. This work is supported by the National Science Foundation under
grant number PHY-2112680, and PHY-2412875. 
\end{acknowledgments}

\bibliography{lepto}

\onecolumngrid

\appendix

\section{Parameters for Leptogenesis Benchmark Points}\label{app:bm_points}

\begin{table*}[htb!]
    \centering
    \resizebox{0.9\textwidth}{!}{\begin{tabular}{|c|c c c||c c c|}
        \hline
        \textbf{Parameter} & \textbf{BM 1 (NO)} & \textbf{BM 2 (NO)} & \textbf{BM 3 (NO)} & \textbf{BM 1 (IO)} & \textbf{BM 2 (IO)} & \textbf{BM 3 (IO)} \\
        \hline
        \hline
        $m_{1(3)}$ [eV] & $10^{-9.7}$ & $10^{-6.1}$ & $10^{-9.9}$ & $10^{-10.0}$ & $10^{-9.5}$ & $10^{-6.7}$ \\
        \hline
        $M_2$ [TeV] & 7.5 & 15 & 40 & 5 & 10 & 25 \\
        \hline
        $\frac{\Delta M_{32}}{M_2}$ & 0.1\% & 0.5\% & 1\% & 1\% & 4\% & 10\% \\
        \hline
        $x_1$ [$^\circ$] & 0.20 & -179.67 & 0.67 & 178.19 & 174.31 & -175.30 \\
        \hline
        $y_1$ [$^\circ$] & -0.42 & -0.01 & 0.29 & 7.70 & -2.40 & 6.09 \\
        \hline
        $x_2$ [$^\circ$] & 170.12 & -47.34 & -169.01 & 102.42 & -23.82 & -127.97 \\
        \hline
        $y_2$ [$^\circ$] & 157.07 & -157.14 & -153.86 & 153.98 & -167.83 & 153.64 \\
        \hline
        $x_3$ [$^\circ$] & -87.54 & -88.67 & 96.87 & 150.66 & -164.83 & 42.02 \\
        \hline
        $y_3$ [$^\circ$] & 3.94 & 3.12 & -1.96 & -159.62 & -178.51 & 170.06 \\
        \hline
        $\delta$ [$^\circ$] & 188.13 & 139.94 & 316.10 & 345.67 & 218.56 & 203.28 \\
        \hline
        $\alpha_{21} [^\circ]$ & 33.13 & 234.96 & 214.07 & 221.49 & 168.83 & 491.63 \\
        \hline
        $\alpha_{31} [^\circ]$ & 5.35 & 602.27 & 466.95 & 262.38 & 348.94 & 451.62 \\
        \hline
        $\theta_{23} [^\circ]$ & 48.57 & 48.51 & 47.93 & 48.83 & 48.45 & 48.36 \\
        \hline
        $\theta_{12} [^\circ]$ & 34.13 & 33.49 & 33.71 & 34.20 & 33.60 & 33.95 \\
        \hline
        $\theta_{13} [^\circ]$ & 8.59 & 8.46 & 8.57 & 8.66 & 8.52 & 8.58 \\
        \hline
        $\tan\beta$ & 200 & 180 & 150 & 225 & 200 & 150 \\
        \hline
    \end{tabular}}
    \caption{Detailed parameters for leptogenesis BMs used in main text in either normal ordering or inverted ordering neutrino spectrum. $M_1 = 1$ GeV are fixed for all BMs.}
    \label{tab:BM_points}
\end{table*}

\end{document}